\documentclass[prl,twocolumn,aps]{revtex4}
\usepackage{amsmath}
\usepackage{amssymb}
\usepackage{graphicx}

\begin{document}
\title{$U(1)$ gauge symmetry free of redundancy and a generalized
 Byers-Yang theorem 
}
\author{Kicheon Kang}
\email{kicheon.kang@gmail.com}
\affiliation{Department of Physics, Chonnam National University, Gwangju 61186, 
 Republic of Korea}

\begin{abstract}
We present a reformulation of the $U(1)$ gauge theory by eliminating the
redundancy inherent in the conventional approach. Our reformulation is
constructed on the basis of local field interaction approach to electrodynamics.
The gauge symmetry in our framework is associated with a physical
transformation, which represents the invariance of 
the equation of motion of a charged scalar field under 
the change in the distribution of electromagnetic field 
at a distance. We demonstrate that all physical properties
of the $U(1)$ gauge theory are preserved with the removal of 
redundancy in the gauge field. In addition, our reformulation provides
a generalization of the Byers-Yang theorem to open systems.  
\end{abstract}

\maketitle

{\em Introduction-}.
Gauge theory is one of the greatest pillars in modern  physics.
It provides a universal framework to understand a wide range of phenomena 
ranging from the field theories of electromagnetism to the standard model
of elementary particles and forces.
Despite its great success, gauge theory consists of a disturbing feature; 
it is constructed based on redundancy of description
rather than the physical symmetry~(see e.g., p.189 of Ref.~\onlinecite{zee10}).
The simplest example is classical electrodynamics.
A point charge $e$ with four-velocity $\dot{r}_\mu$ 
under the four-potential $A^\mu$ is described by the Lagrangian:
\begin{equation}
 L = L_0 + \frac{e}{c}\dot{r}_\mu A^\mu ,
\label{eq:Lin_A}
\end{equation}
where $L_0$ is the kinetic part of the particle. 
Gauge symmetry in this Lagrangian implies the invariance of the
equation of motion for the particle under transformation
\begin{equation}
 A^\mu \rightarrow A'^\mu = A^\mu - \partial^\mu\Lambda 
\label{eq:gauge-tr}
\end{equation}
with any scalar function $\Lambda(x)$. 
Notably, this transformation is not associated with the symmetry 
of two physical states 
that have the same properties. Instead, it indicates that $A^\mu$ and $A'^\mu$
represent the same physical state.
In other words, it is impossible, even in principle,
to make a gauge transformation of the system in a laboratory, 
unlike other physical 
transformations, e.g., translation, rotation, and the Lorentz transformation.
 
This property of gauge theory is closely related to the ``nonlocality"
of electromagnetic interaction because the local interaction of the particle
with the gauge field (Eq.~\eqref{eq:Lin_A}) includes a
certain degree of arbitrariness (see Eq.~\eqref{eq:gauge-tr}). 
Recently, it was found that this arbitrariness of the
interaction can be removed by adopting the local field interaction (LFI)
approach~\cite{kang13,kang15,kang17,kim18}. 
The LFI theory successfully reproduces the classical electrodynamics
and topological 
Aharonov-Bohm (AB) effect~\cite{kang13,kang15}. 
This implies that we can describe quantum theory involving electromagnetic
interaction that does not rely on the potential, 
in contrast to the conventional viewpoint.
However, one may inquire whether 
the approach without redundancy of $A^\mu$ removes 
the ubiquity of gauge theory.
We need to clarify the manifestation of ``gauge symmetry"
in this redundancy-free description. 
In this Letter, we resolve this question 
and reveal the gauge symmetry intrinsic in the LFI approach. 
The gauge symmetry in this
framework is associated with a physical transformation without mathematical 
redundancy, while preserving other properties of $U(1)$ gauge theory.
We discuss the gauge symmetry in this framework 
for both a point charge and charged scalar field.
In addition, applying this gauge symmetry to a system
without a closed loop, we derive a generalized 
Byers-Yang theorem~\cite{byers61,imry02} and show that it can be 
experimentally verified in a superconducting point contact.

{\em Gauge invariance as a physical symmetry in classical electrodynamics-}.
A particle with charge $e$ and mass $m$ under external electric ($\mathbf{E}$)
and magnetic~($\mathbf{B}$) fields can be described
by the Lagrangian:
\begin{equation}
 L = L_0 + L_{in} ,
\label{eq:L}
\end{equation}
where
\begin{equation}
 L_0 = -mc\sqrt{c^2-\dot{\mathbf{r}}\cdot\dot{\mathbf{r}}}
\label{eq:L0}
\end{equation}
is the kinetic part, and 
\begin{subequations}
 \label{eq:Lin_Pi}
\begin{equation}
 L_{in} = \dot{r}_\mu \Pi^\mu 
\end{equation}
represents the interaction of a particle with the external fields 
given in terms of the Lorentz-covariant LFI~\cite{kang13,kang15}.
In this description, redundancy of the potential $A_\mu$ 
in Eq.~\eqref{eq:Lin_A} is eliminated and the particle's motion is 
coupled with the ``field-momentum four vector", 
$\Pi^\mu=(\Pi^0,\mathbf{\Pi})$,
defined as
\begin{eqnarray}
 & & \Pi^0 = \frac{1}{4\pi c}\int \mathbf{E}_e\cdot\mathbf{E}\, d^3\mathbf{x}, 
  \\
 & & \mathbf{\Pi} = \frac{1}{4\pi c}\int \mathbf{E}_e\times\mathbf{B}\, 
 d^3\mathbf{x}, 
\end{eqnarray}
\end{subequations}
where $\mathbf{E}_e$ is the electric field generated by charge $e$.
Mathematically, $\Pi^\mu$ in Eq.~\eqref{eq:Lin_Pi} plays the same
role as $A^\mu$ in the potential-based Lagrangian in Eq.~\eqref{eq:Lin_A}.
This Lagrangian successfully reproduces the classical
equation of motion and the topological quantum phase. 
In addition, it demonstrates the locality of the interaction which 
can be tested in real experiments.
For its details, see Refs.~\onlinecite{kang13,kang15,kang17,kim18}.

We aim to demonstrate the appearance of gauge symmetry in the absence of
redundancy in the gauge field.
For a given system configuration, the Lagrangian of Eq.~\eqref{eq:L}
is unique; if $\Pi^\mu$ is different in $L_{in}$, 
it represents a different
configuration of the external $\mathbf{E}$ and $\mathbf{B}$. 
The Lagrangian possesses a symmetry under this condition.
For the transformation 
\begin{equation}
 \Pi^\mu \rightarrow \Pi'^\mu = \Pi^\mu - \partial^\mu\Lambda \,
\label{eq:gauge-tr-Pi}
\end{equation}
the Lagrangian is transformed as
\begin{equation}
 L \rightarrow L' = L - \frac{d\Lambda}{dt} \,.
\end{equation}
This indicates that the equation of motion for the particle is
invariant under the ``gauge transformation" of Eq.~\eqref{eq:gauge-tr-Pi}
because a total time derivative $d\Lambda/dt$ does not affect the Lagrange
equation of motion~(see e.g., Section 2 of Ref.~\onlinecite{landau60}). 
From the Lagrangians \eqref{eq:L}, \eqref{eq:L0}, and \eqref{eq:Lin_Pi}, 
we obtain the gauge-invariant equation of motion:
\begin{subequations}
\label{eq:eom}
\begin{equation}
 \frac{dp^\mu}{dt} = G^{\mu\nu} \frac{dr_\nu}{dt} \,, 
\end{equation}
where $p^\mu$ is the momentum 4-vector of the particle and
\begin{equation}
 G^{\mu\nu} \equiv \partial^\mu\Pi^\nu-\partial^\nu\Pi^\mu 
    = \frac{e}{c}F^{\mu\nu}
 \,,  
\end{equation}
\end{subequations}
is proportional to the electromagnetic field tensor $F^{\mu\nu}$. 
Eq.~\eqref{eq:eom}
is equivalent to the one obtained from the standard potential-based
approach~(see e.g., Ref.~\onlinecite{jackson99})
and can be rewritten as
\begin{equation}
 \frac{d\cal E}{dt} = e\dot{\mathbf{r}}\cdot\mathbf{E}, \;\;\;
 \frac{d\mathbf{p}}{dt} = e\mathbf{E} 
    + \frac{e}{c}\dot{\mathbf{r}}\times\mathbf{B} \,,
\end{equation}
where ${\cal E}, \mathbf{p}$, and  $\dot{\mathbf{r}}$ are 
the energy, momentum, and velocity of the particle, respectively.

The invariance of the equation of motion 
under the gauge transformation of $\Pi^\mu$ 
(Eq.~\eqref{eq:gauge-tr-Pi}) is not associated with the
redundancy of description because $\Pi^\mu$ is a physical quantity 
without arbitrariness.
The change in $\Pi^\mu$ in Eq.~\eqref{eq:gauge-tr-Pi}
is caused by the variations in $\mathbf{E}$ and $\mathbf{B}$. 
A constraint in this
transformation is that the local value of field tensor
$F^{\mu\nu}$ remains unchanged. In other words,
the ``gauge transformation" of $\Pi^\mu$
represents a change in 
distribution in the external $\mathbf{E}$ and $\mathbf{B}$, while
their {\em local} values remain unchanged
at the position of the particle. Therefore, the gauge symmetry in the Lagrangian
for a point charge (Eqs.~\eqref{eq:L}-\eqref{eq:Lin_Pi}) implies that
{\em its equation of motion is invariant under any change in the external field
distribution at a distance.} 
This invariance is evident because the equation of motion is local, 
although the conventional gauge
theory with $A^\mu$ does not consider the problem in this way. 

{\em $U(1)$ gauge invariance as a physical symmetry-}.
Let us consider electrodynamics with charged scalar field $\phi$.
Applying the canonical transformation of the Lagrangian 
(Eqs.~\eqref{eq:L}-\eqref{eq:Lin_Pi}) with the introduction of
$\phi$, we obtain the Klein-Gordon equation:
\begin{equation}
 \left[ 
  -(\partial_\mu-\frac{i}{\hbar}\Pi_\mu)(\partial^\mu-\frac{i}{\hbar}\Pi^\mu)
  + \frac{m^2c^2}{\hbar^2} \right] \phi = 0.  
\label{eq:KG}
\end{equation}
It should be noted that we may work on the Dirac field for the electron, but
it gives the same result for the $U(1)$ gauge symmetry.
The Klein-Gordon equation~(Eq.~\eqref{eq:KG}) for $\phi$ is 
generated by the Lagrangian:
\begin{eqnarray}
 {\cal L} &=& -\frac{1}{m} 
 (\hbar\partial_\mu\phi-i\Pi_\mu\phi) (\hbar\partial^\mu\phi^*+i\Pi^\mu\phi^*)  
    \nonumber - mc^2\phi^*\phi 
  \\ & & - \frac{1}{16\pi} F_{\mu\nu}F^{\mu\nu} \,.
\label{eq:L-KG}
\end{eqnarray}

The mathematical structure of this Lagrangian is equivalent
to that given by the standard approach where $\Pi_\mu$ 
is replaced by $eA_\mu/c$. 
Therefore, exploring the $U(1)$ gauge symmetry
is straightforward and it produces the following results.
First, Lagrange equations for 
the fields $\phi$ and $\Pi_\mu$ lead to Klein-Gordon~\eqref{eq:KG}
and Maxwell equations, respectively. 
Second, and most importantly, the gauge symmetry 
is manifested in the invariance of ${\cal L}$ under the transformation
\begin{equation}
 \phi\rightarrow \phi' = \phi e^{-i\Lambda/\hbar}, \;\;\;
 \Pi_\mu\rightarrow \Pi_\mu' = \Pi_\mu - \partial_\mu\Lambda ,
\end{equation}
with an arbitrary scalar function $\Lambda$. 
Similar to the case of the point particle discussed above, 
this transformation does not include any redundancy of description. 
It is a physical
symmetry associated with different $\Pi_\mu$, or equivalently, with different
distributions of external $\mathbf{E}$
and $\mathbf{B}$. The gauge symmetry in ${\cal L}$ indicates that
{\em the equation of motion for $\phi$ {\rm (Eq.~\eqref{eq:KG})} is invariant 
under the change of the external electromagnetic field at a distance.}
Finally, the charge conservation is derived from the symmetry via the
N\"other's theorem. In our framework, 
it is expressed in terms of continuity equation:
\begin{equation}
 \partial_\mu j^\mu = 0 \,,
\end{equation}
for the four-charge current, 
\begin{equation}
 j^\mu = -i\frac{\hbar}{m}(\phi^*D^\mu\phi - \phi D^\mu\phi^*) \,,
\end{equation}
where the covariant derivative is given by
\begin{equation}
 D^\mu\phi = (\partial^\mu - \frac{i}{\hbar}\Pi^\mu)\phi \,.
\end{equation}

{\em Generalized Byers-Yang theorem-}.
An intriguing consequence of the $U(1)$ gauge symmetry is the 
Byers-Yang theorem~\cite{byers61}. It states that all physical properties of
a doubly connected system (an annulus) enclosing a magnetic flux $\Phi$ 
(see Fig.~1(a)) are periodic in $\Phi$ with period $\Phi_0=hc/e$. 
Here, we show that the theorem can be extended to an open system (Fig.~1(b))
with our formulation.
We also propose an experimental arrangement to confirm the generalized 
theorem using the superconducting point contact. 
In our framework of the LFI approach, the eigenfunction $\phi$ of 
a charged particle in both systems (Fig.~1(a,b)) satisfies the wave equation
\begin{equation}
 \frac{1}{2m}(-i\hbar\nabla - \mathbf{\Pi})^2\phi + c\Pi^0\phi = \epsilon\phi 
  \,,
\end{equation}
in the limit $v/c\ll 1$ of Eq.~\eqref{eq:KG}.
For many particles, the energy eigenfunction $\psi$ satisfies
\begin{equation}
 \frac{1}{2m}\sum_j 
  (-i\hbar\nabla_j-\mathbf{\Pi}(\mathbf{r}_j))^2 \psi 
      + V\psi = E\psi  \,.
\label{eq:wave-psi}
\end{equation}
The magnetic field vanishes in the region of nonzero $\psi$, 
 $\mathbf{B} = (c/e) \nabla\times \mathbf{\Pi} = 0$.
Therefore we can write $\mathbf{\Pi} = \nabla\Lambda$, 
and $\mathbf{\Pi}$ can be gauged away. 
Under the transformation 
\begin{eqnarray}
 \psi &\rightarrow& \psi' = 
   \psi e^{-(i/\hbar)\sum_j\Lambda(\mathbf{r}_j)},  
   \nonumber \\
 \mathbf{\Pi} &\rightarrow& \mathbf{\Pi}' = \mathbf{\Pi} - \nabla{\Lambda} = 0
 \,,
\label{eq:gauge-tr-pure}
\end{eqnarray}
the wave equation~\eqref{eq:wave-psi} reduces to
\begin{equation}
 \frac{1}{2m}\sum_j 
  (-i\hbar\nabla_j)^2 \psi' 
      + V\psi' = E\psi'  \,,
\label{eq:wave-psi'}
\end{equation}
implying that $\mathbf{\Pi}$ is removed from the wave equation 
with a modified boundary condition in $\psi'$.

Consider the boundary condition of a doubly connected system. 
For any specific coordinates of a particle, say $\mathbf{r}_i$, 
that circulates around the loop once
while keeping the other coordinates fixed, $\psi'$ acquires
a phase factor by the transformation \eqref{eq:gauge-tr-pure} as
\begin{equation}
 \psi' \rightarrow \psi' e^{-i\oint\mathbf{\Pi}\cdot d\mathbf{r}/\hbar} 
   = \psi' e^{-i(e\Phi/\hbar c)} \,. 
\label{eq:bc}
\end{equation}
Because the wave equation~\eqref{eq:wave-psi'} is independent of 
$\mathbf{\Pi}$, the $\mathbf{\Pi}$-dependence of $E$
is determined by the boundary condition \eqref{eq:bc}, which constitutes the 
original Byers-Yang theorem: all physical properties of the loop 
are periodic in $\Phi$ with its period $\Phi_0 = hc/e$. 

Our analysis on the periodicity can also be applied to an open system 
(Fig.~1(b)).
For a specific coordinate of a particle, $\mathbf{r}_i$, let
$\psi'_L$($\psi'_R$) be the asymptotic value of the wave function 
at the left (right) infinity of $\mathbf{r}_i$ such that
\begin{equation}
\psi'_L \equiv \psi'(\mathbf{r}_i\rightarrow -\infty), \;\;\;
\psi'_R \equiv \psi'(\mathbf{r}_i\rightarrow \infty). 
\end{equation}
From the gauge transformation (Eq.~\eqref{eq:gauge-tr-pure}), we obtain 
\begin{equation}
 \frac{\psi'_L}{\psi'_R} = \alpha
   e^{i\int_{-\infty}^\infty \mathbf{\Pi}\cdot d\mathbf{r}/\hbar } ,  
\label{eq:bc_open}
\end{equation}
where the constant $\alpha$ is independent of $\mathbf{\Pi}$.
Because the eigenvalue $E$ is determined by
the wave equation \eqref{eq:wave-psi'} and the boundary condition
\eqref{eq:bc_open}, all physical properties are periodic in 
 $\int_{-\infty}^\infty \mathbf{\Pi}\cdot d\mathbf{r}/\hbar$
with period $2\pi$.
This indicates that the Byers-Yang theorem can be extended to an open system. 

Before discussing a realistic example that demonstrates 
the Byers-Yang theorem for an open system,
let us point out some important facts. 
First, there are no observable effects of the external flux 
in a normal conductor in
the configuration of Fig.~1(b), because
the boundary condition does not alter the physics of the open system. 
However, the situation is different for a superconductor with
macroscopic quantum coherence. This is analyzed below in detail. 
Second, the standard approach with vector potential $\mathbf{A}$ 
fails to describe
the periodicity in the open system. In the potential-based approach, the
boundary condition of Eq.~\eqref{eq:bc_open} is replaced by
\begin{equation}
 \frac{\psi'_L}{\psi'_R} = \alpha
   e^{i(e/\hbar c)\int_{-\infty}^\infty \mathbf{A}\cdot d\mathbf{r} } ,
\end{equation}
and its phase factor remains ambiguous as the integral 
$\int_{-\infty}^\infty \mathbf{A}\cdot d\mathbf{r}$ is not a well-defined
quantity for an open path.

{\em Andreev bound states and gauge symmetry-}.
Now we discuss the manifestation of the Byers-Yang theorem for an open system 
with the boundary condition of Eq.~\eqref{eq:bc_open}
in a realistic system. The system under consideration 
is a Josephson weak link that connects the two regions of a superconductor 
with an external magnetic flux at a distance of the superconductor (see Fig.~2).
The type of the junction is insignificant here.
It can be described by the Bogoliubov-deGennes equation~\cite{degennes66}: 
\begin{subequations}
\begin{equation}
 \left( \begin{array}{cc} 
         H_e & \Delta(x) \\ 
         \Delta^*(x) & -H_e^*
        \end{array} \right)
 \left( \begin{array}{c}
         u(x) \\ v(x) 
        \end{array} \right) 
 = E 
 \left( \begin{array}{c}
         u(x) \\ v(x)
        \end{array} \right) ,
\end{equation}
where the components of the Hamiltonian in our framework is given by
\begin{eqnarray}
 H_e &=& \frac{1}{2m}(-i\hbar\nabla-\mathbf{\Pi})^2 + U(x) , \nonumber \\
 H_e^* &=& \frac{1}{2m}(i\hbar\nabla-\mathbf{\Pi})^2 + U(x) . 
\end{eqnarray}
\end{subequations}
$\mathbf{\Pi}$ can be gauged away by the following transformation 
to the `primed' functions:
\begin{eqnarray}
 \mathbf{\Pi}' &=& \mathbf{\Pi} - \nabla\Lambda = 0 \, \nonumber \\
 u' &=& ue^{-i\Lambda/\hbar}, \;\; v'= ve^{i\Lambda/\hbar}, 
 \label{eq:gauge-tr-sc}\\
 \Delta' &=& \Delta e^{-2i\Lambda/\hbar}, \nonumber
\end{eqnarray}
and thus, we obtain 
\begin{subequations}
\label{eq:bdGe'}
\begin{equation}
 \left( \begin{array}{cc} 
         H_e' & \Delta'(x) \\ 
         \Delta'^*(x) & -H_e'
        \end{array} \right)
 \left( \begin{array}{c}
         u'(x) \\ v'(x) 
        \end{array} \right) 
 = E 
 \left( \begin{array}{c}
         u'(x) \\ v'(x)
        \end{array} \right) ,
\end{equation}
where 
\begin{equation}
 H_e' = -\frac{\hbar^2}{2m}\nabla^2 + U(x) \,.
\end{equation}
\end{subequations}
This transformation reveals the periodicity of the physical properties of the 
system.
The eigenvalue $E$ is determined by Eq.~\eqref{eq:bdGe'} and
the boundary condition
of $\Delta'$ (represented in its phase shift)
\begin{subequations}
\label{eq:varphi}
\begin{equation}
 \varphi \equiv \arg{(\Delta'_L/\Delta'_R)} = \varphi_0 + \varphi_B ,
\label{eq:varphi_0}
\end{equation}
where
$\Delta'_L\equiv\Delta'(x\rightarrow-\infty)$ and
$\Delta'_R\equiv\Delta'(x\rightarrow\infty)$ are the boundary values
of $\Delta'(x)$ at each lead. 
$\varphi_0 = \varphi_L-\varphi_R$ is the intrinsic
phase difference between the two sides of the superconductor,
and $\varphi_B$ is the flux dependence of the phase given by
\begin{equation}
 \varphi_B = 
  \frac{2}{\hbar} \int_{-\infty}^\infty \mathbf{\Pi}\cdot d\mathbf{r}  
  = \frac{2e\theta}{hc}\Phi \,, 
\end{equation}
\end{subequations}
where $\theta$ is the angle formed in the geometry of the system (see Fig.~2).
Therefore, the eigenvalues are periodic functions of $\Phi$ with period 
$2\pi hc/(2e\theta)$ and all physical properties display the
same periodicity.
Notably, for $\theta=2\pi$, $\varphi_B$ reduces to the Aharonov-Bohm 
phase $2e\Phi/(\hbar c)$ 
associated with the Cooper pair charge $2e$. 

As an example, we consider a delta-function potential 
$U(x) = U_0 \delta(x)$ 
and a constant gap function $\Delta(x) = \Delta_0$.
The latter condition gives $\varphi_0=0$ in Eq.~\eqref{eq:varphi_0}.
A solution inside the gap ($-\Delta_0 < E < \Delta_0$), 
known as the Andreev bound state,
can be determined by solving 
the Bogoliubov-deGennes equation \eqref{eq:bdGe'} with the boundary
condition of $\Delta'(x)$ 
(Eq.~\eqref{eq:varphi})~\cite{kulik69,furusaki99}. We obtain
\begin{equation}
 E = \pm\Delta_0 \sqrt{ 1 - T\sin^2{(\varphi_B/2)} } ,
\label{eq:abs}
\end{equation}
where $T=1/(1+Z^2)$ is the transmission probability across the point contact
with the parameter $Z\equiv mU_0/(\hbar^2 k_F)$ ($k_F$ being 
the Fermi wave vector).
Considering that the Andreev bound states and their phase dependence 
have been well
confirmed in experiments with superconducting hybrid junctions, 
the flux dependence of the bound-state energy
(Eq.~\eqref{eq:abs}) can also be observed in real experiments. 
The bound-state energy may be directly probed by
spectroscopic measurements~(see e.g., Ref.~\onlinecite{pillet10,kim13})
with variations in the magnetic flux.
To confirm the generalization of the Byers-Yang theorem,  
the superconductor should not form a
closed loop that circulates around the flux to avoid observation of the ordinary
AB phase $2e\Phi/\hbar c$.

{\em Conclusion-.} 
In conclusion, we have presented a reformulated $U(1)$ gauge theory 
on the basis of physical symmetry.
The symmetry transformation corresponds to a change in the electromagnetic
field in the inaccessible region of the charged scalar field ($\phi$)
along with a change in the phase factor of $\phi$.
This reformulation preserves all properties of the $U(1)$
gauge theory but eliminates the redundancy inherent in the
conventional approach. This also implies that 
quantum electrodynamics can be defined without relying on $A^\mu$.
In addition, our formulation provides a generalization of 
the Byers-Yang theorem to an open system, 
which can be confirmed in an experiment
for the Andreev bound states of a superconducting point contact.

 \bibliography{references}

\newpage
\begin{figure}
\centering
\includegraphics[width=8cm]{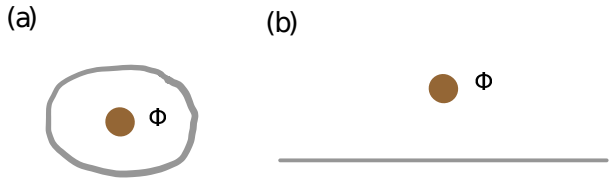} 
\caption{(a) Doubly connected system of conductor (gray region) with
external magnetic flux $\Phi$ pierced inside closed loop. 
(b) Similar, but open, conductor with external $\Phi$.
In both systems, gauge symmetry provides periodicity of 
energy eigenvalues as a function of $\Phi$.
 }
\end{figure}
\begin{figure}
\centering
\includegraphics[width=6cm]{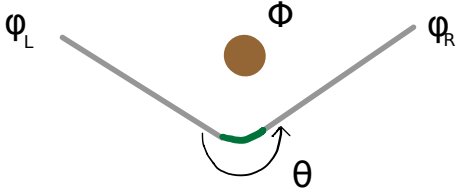}
\caption{Superconducting point contact with external magnetic flux $\Phi$.
 Andreev bound state energies depend on the phase difference 
 $\varphi_L-\varphi_R$ between two superconductors, $\Phi$, and angle
 $\theta$ formed in the geometry of the system.
 }
\end{figure}
\end{document}